\documentclass{aa}
\usepackage{graphicx}
\begin{document}

\def\Q1429{LBQS~1429-0053}
\def\hyp{Hyperz }   
\def\ho{H$_0$ }
\def\hirac{HiRAC }
\def\sex{SExtractor }
\def\kmsmpc{km~s$^{-1}$~Mpc$^{-1}$}

   \title{LBQS~1429-0053: a binary quasar rather than a lensed quasar
\thanks{Based on observations
obtained with VLT/ANTU at  ESO-Paranal Observatory (program
67.A-0502) and with the Hubble Space Telescope, operated by NASA.}}
    
\subtitle{}

\author{C. Faure\inst{1,3}, D. Alloin\inst{1}, S. Gras\inst{1}, 
F. Courbin\inst{2}, J.-P. Kneib\inst{3,4},  P. Hudelot\inst{3}} 
   
   \offprints{C\'ecile Faure: cfaure@eso.org}

   \institute{ European Southern  Observatory, Alonso de Cordova 3107,
         Casilla   19001,    Santiago   19,   Chile    
         \and   
         Institut
         d'Astrophysique et de  G\'eophysique, Facult\'e des Sciences,
         Universit\'e de Li\`ege, All\'ee du 6 ao\^ut, 17, B5C, Li\`ege 1,  
         Belgium 
         \and  
         Observatoire  Midi$-$Pyr\'en\'ees, UMR
         5572, 14  avenue Edouard  Belin, 31400 Toulouse,  France 
         \and
         Caltech Astronomy  Department, Mail Code  105-24, Pasadena, CA
         91125, USA}

\date{Submitted:17/12/2002}

\authorrunning{C. Faure et al.}
\titlerunning{LBQS~1429-0053: a binary quasar rather than a lensed quasar}

\abstract{Very deep ESO/VLT FORS1 and  ISAAC images, as well as HST
NICMOS2  data are  used to  infer the  nature of the  quasar pair
\Q1429  A and  B, either  a binary  quasar or  a  doubly-imaged lensed
quasar.  Direct  search of a  putative lensing galaxy  is unsuccessful
down  to  R=27,  J=24,  Ks=22.5  and
H=22.5. Moreover, no  galaxy overdensity close to the quasar pair  is found. A weak shear analysis of the FORS1 $R$-band 6.8\arcmin\,$\times$6.8\arcmin\,
field also  fails at detecting any  concentration of dark
matter more massive than $\sigma$=500 km~s$^{-1}$ and  weakens the hypothesis  of a  dark lens.  The  only sign  of a
possible lens consists in a group  of  5 objects having  colors consistent
with galaxies at z$\sim$1, within  a radius of 5\arcsec\, from the
quasar pair. Considering this  group as the lensing potential  does not allow to
reproduce  the image  position  and  flux ratio of  \Q1429 A and B. Our  deep
$R$-band image  shows a blue, previously unknown, extended  object at
the  position of  \Q1429 A, which  is  consistent  with either being the
lensed quasar A  host, or being an intervening galaxy at lower redshift. Unless future
very deep optical images demonstrate that this object is actually the lensed
host of \Q1429, we conclude that there is very little evidence for
\Q1429 being lensed. Therefore, we are led to declare \Q1429 A and B a genuine binary quasar.
 }

 \maketitle
%
\section{Introduction}

Doubly-imaged quasars  were initially thought to be  the easiest cases
of gravitational  lenses to  interpret and model.   The first discovered
gravitational lens  was indeed  a double, Q~0957+561  (Walsh, Carswell
and Weymann, 1979), with an  image separation of 6\arcsec. Yet, doubles
are  just  as  challenging   to  interpret  as  more  complex  image
configurations.  In addition,  while the  lensing  interpretation is
often  invoked to  explain quasar pairs, they  could  well  be genuine
binary quasars. As  a matter of
fact, about  14 quasar pairs  with large angular  separation (larger
than 3 \arcsec) remain inconclusive  cases to whether they are binary
quasars or doubly-imaged lensed quasars. The quasar pair \Q1429 is one
of them (Hewett et al. 1989).

The long-standing  question whether some of the  observed quasar pairs
are doubly-imaged  lensed quasars or  genuine binary quasars  has been
revived  through recent  discussions by  e.g.  Mortlock  et al.(1999),
Kochanek  et al. (1999)  and Peng  et al  (1999). This  question bears
important cosmological implications. If  an observed pair is a genuine
binary quasar,  the spectral  similarity of the  two components  is an
exciting feature telling us about the formation and feeding of massive
black  holes.  If  it  is a  doubly-imaged  quasar, the  absence of  a
detectable luminous lens is pushing forward the idea of dark lenses. 
Understanding the nature of double quasars is therefore of great importance.

Recently, Peng et al. (1999)  have touched upon this problem, studying
the  nature of  the quasar  pair Q~1634+267.  They do  not  detect any
lensing galaxy, but, because of the close similarity of the quasar spectra, they still envisage  the dark  lens  hypothesis.  To  confirm their hypothesis,  Peng et al. (1999) suggest to measure the light curves of both quasars, as they should be almost  identical in  case of gravitational lensing.  Another
complementary way is to try to directly detect a large mass concentration, even dark, along the line of sight, by performing a  weak lensing analysis or deep X-ray observations.

The quasar pair \Q1429 A and B (component A: $\alpha$=14:32:29.3, $\delta$=-01:06:16, J2000)
 was  first  observed  in  
spectroscopic  mode at  the MMT  (Hewett et  al.   1989). The
spectra of \Q1429 A and \Q1429 B were found to be very similar, showing
the  emission lines  and blends  of Ly$\alpha$-NV,  SiIV-OIV,  CIV and
CIII],  very common  in  quasar spectra. An additional spectrum of \Q1429 A reveals strong MgII emission line (Mortlock et al. 1999). The  quasar redshift is
$z=2.076$.   The  separation of  the  two  components  is  large,
5.1\arcsec.  A  search for  a putative lensing  object, on  an image
obtained at CFHT down to R=24.5, remained unsuccessful. But this does not rule  out the presence of a lens (see for example the case of the Cloverleaf, Kneib et al. 1998).

  \Q1429  has  been  classified as  an  O$^2$ pair  by
Kochanek  et   al.  (1999),  meaning  that  both   quasars  are  radio
faint. According to the statistical analysis presented in that paper, the    O$^2$ property does not allow to conclude whether the quasar pair is a lensed system or a genuine binary quasar. 
Moreover, in the analysis of Mortlock et  al. (1999), the nature of \Q1429 could
not  be established either, with any  certainty: according  to these  authors, the
degree of   similarity between the   spectra  from  \Q1429 A  and  \Q1429 B,  is
consistent with this system being either a lensed doubly-imaged quasar
or  a binary  quasar.

   If  \Q1429 is
actually lensed, the spectral  differences between A and B could still be attributed to a number of effects: (a) to microlensing by stars in  the lensing galaxy, (b) to absorption by dust
in the  lens, or (c) to  the long time-delay expected  for the system
and implying  that images A and B could correspond to the  quasar seen  at two different epochs,
hence possibly in a different state of  activity.  A tentative model
of the  lensing object was  proposed by Hewett  et al (1989),  but was
largely under-constrained. Assuming that \Q1429  is lensed, two models were
considered.  The first model had  a deflector (point mass) of cluster mass ($\mathcal{M}\sim$1~10$^{13}\mathcal{M}\odot$), at the redshift of one of the two observed MgII absorption system, z$_l\sim$1.5. The second  model, with z$_l\sim$0.6,  implied a
deflector with an extremely  high mass-to-light ratio ($\mathcal{M}\sim$2~10$^{13}\mathcal{M}\odot$ and $\mathcal{M}/\mathcal{L}_{R}\sim$3000), in other words,
a ``dark lens''.

The  availability of  large  telescopes allows  to  push further  the
direct search for  a putative lens in the vicinity  of the two quasar
components. Alternatively,  if there were  a massive dark lens  in the
field of \Q1429  it would produce a weak  but measurable distortion on
all background  galaxies. Looking for a  weak lensing effect  is another
way of revealing  a dark lens, using wide-field  deep observations. This method  has been used and has led to  positive detections on  wide field images
of  Q2345+007  (Bonnet et  al.  1993) and  MG 2016+112  (Benitez  et
al. 1999). The  cluster responsible for the weak  lensing signal in MG
2016+112  has recently  been  confirmed    spectroscopically  (Soucail  et
al. 2001).

We have  observed \Q1429 in  the course of  a larger project  aimed at
analyzing in  detail the mass  distribution towards a large  number of
known multiply-imaged quasars and  towards some suspected lenses.  The
ESO  8.2 m VLT allowed to  reach 
limiting magnitudes  of R=27.0, J=24.0  and Ks=22.5 (3$\sigma$),
in conditions of very good image quality, for most of our targets.

In Sect. 2, we discuss the VLT and HST observations and the  data reduction.
In  Sect. 3  we  present  the  photometric properties  of the  quasar pair  
\Q1429 A and  B and  of the  closest
galaxies, as inferred either from PSF subtracted or deconvolved
images in the $R$-, $J$-, $H$- and $Ks$-bands. We discuss in Sect. 4 the
galaxies close to the quasar components and the possible detection of
the  quasar host  galaxy.  Two  different methods  to  look for  the
putative  lensing potential  are discussed  in Sect.  5.  Finally,  two tentative lens models are discussed, while the
results  and  conclusions of the paper are summarized  in  Sect. 6. 
We adopt $H_0$=65 \kmsmpc, $\Omega$=0.3 and $\Lambda$=0.7
throughout.

\begin{table}[t!]
\renewcommand{\arraystretch}{1.0}
\centering
\begin{center}
\caption[]{Summary  of  the instrumental  configuration  used for  the
optical and near-IR observations  of \Q1429. The columns display respectively the date
of observation, the telescope and instrument used, the 
filter, the image
quality (IQ), and the total exposure time in kilo-seconds.  }
\begin{tabular}{lllll}
\hline
Date           & Instrument & Filter & IQ & Time  \\
\hline\hline
2001~Jun~08,15 & VLT-ISAAC  & $Ks$    & 0.64\arcsec     & 3.7\\
2001~Jun~08    & VLT-ISAAC  & $J$      & 0.50\arcsec    & 2.6\\
2001~Apr~18    & VLT-FORS1  & $R$     &   0.55\arcsec   & 3.6\\
1997~Dec~30    & HST-NICMOS & $H$      & 0.16\arcsec     &  2.6\\ 
\hline
\label{obs}
\end{tabular}
\end{center}
\end{table}

\section{Observations and Data Reduction}

The ground-based observations presented  in this paper were taken with
the  Infrared Spectrometer  And  Array Camera  (ISAAC, Cuby et al. 2002)  and the  Focal
Reducer/low  dispersion  Spectrograph   (FORS1, Szeifert 2002)  attached  to  the  8.2 m
telescope VLT/ANTU,  at ESO/Paranal. The observations  were obtained in
Period 67  as part  of programs  67.A-502(A,B,C).  Observations  with the Hubble  Space Telescope  (HST) and  the near-IR
NICMOS  instrument were  all retrieved from  the HST/STScI  archive.  Detailed information about the  observational setup and conditions are
presented hereafter.

\subsection{VLT Observations}

Direct  near-IR ISAAC  imaging  in the  $J$- and $Ks$-bands has  been
carried out  with the Short Wavelength Imaging  camera (SWI1) equipped
with a Rockwell-Hawaii (RH) 1024 $\times$ 1024 pixel array.  The scale
is 0.1484\arcsec\, per pixel and the field-of-view, 2.5\arcmin $\times$
2.5\arcmin.  The scientific and  calibration frames were corrected for
the odd-even  column effect arising  from an ISAAC  misfunction during
the  nights of  observation:  this artifact  was successfully  removed
through a Fourier analysis procedure  (see ISAAC handbook, Cuby et al.
2002).  Standard  reduction procedures  were then applied  to subtract
the dark  and to flat-field the  images using a  normalized flat.  Sky
subtraction  and  co-addition  of  the reduced  dithered  frames  were
performed using tasks from XDIM/IRAF.

The FORS1 instrument  was used to carry out  the $R$-band observations
using the standard resolution collimator (SR).  The resulting scale is
0.2\arcsec\, per pixel  and  the  useful  field-of-view is  6.8\arcmin\, $\times$  6.8\arcmin.   The  $R$-band  data  were  reduced  using  sky
flat-fields  and  standard  IRAF   routines.   A summary  of the  main observational  parameters  can be
found in Table~\ref{obs}.

\subsection{Hubble Space Telescope observations}

HST archival data of \Q1429 are available as part of a much larger
public survey  of gravitational  lenses (PI: E.   Falco) known  as the
CfA-Arizona Space Telescope  LEns Survey (CASTLES, Kochanek et al. 2002).  The NICMOS2/F160W
($H$-band) data were taken on  1997 December 12, with a total exposure
time  of 4 $\times$  640s.  At  the date  of observation  the NICMOS2
pixel size was 0.0760\arcsec\, $\times$ 0.0753\arcsec.  We constructed a
mosaic   of   the  flux   calibrated   frames   using  standard   IRAF
procedures. The final image is a simple combination of the frames.  In
performing the  image combination, a  mask image has been  created and
used  to remove  bad pixels  and  cosmic rays.  
The  resolution of  the combined image  is $\sim$0.16\arcsec\,
(FWHM).

\section{Subtraction of the quasar components and deconvolution}

In  the first  place, combined images are  used to  search  for the
presence  of  a   single  lens  very  close  to  the  quasar
components.  Even
under  excellent  seeing one  should  apply  numerical techniques  to
extract the signal in an optimum way. We used three techniques: subtraction of the telescope Point-Spread-Function (PSF) and two different image deconvolution techniques in order to 
validate the detection of any faint source.

\begin{figure*}
\begin{center}
\caption{\label{R}a)  The   13\arcsec$\times$13\arcsec  $R$-band  field
around \Q1429 A and B from FORS1.  North is to the top, East is to the
left. b) Same area, but with  the PSF subtracted on components A and B
as explained  in Sect. \ref{sub}.  c) Result of the  MEM deconvolution
(Sect. \ref{deconv}): A and B are the two quasar components. d) Result
of  the  MCS deconvolution  (Sect.  \ref{deconv}).  The photometric
properties of the faint  objects labeled \#1  to 7 are described 
in Sect. \ref{vici} and Table \ref{photometry}. }
\end{center}
\end{figure*}

\begin{figure*}
\begin{center}
\caption{\label{J}a)  The  13\arcsec$\times$13\arcsec $J$-band  field
around \Q1429 from ISAAC. North is to the top, East is to the left. b)
Same  area,  but  with  the  PSF  subtracted on  components  A  and  B
(Sect.    \ref{sub}).   c)   Result    of   the    MEM   deconvolution
(Sect. \ref{deconv}): A and B are the two quasar components. d) Result
of  the  MCS deconvolution  (Sect.  \ref{deconv}).  The photometric
properties of the faint  objects
labeled \#1  to 5 are described in Sect.  \ref{vici} and
Table \ref{photometry}.  }
\end{center}
\end{figure*}

\begin{figure*}
\begin{center}
\caption{\label{K} a) The  13\arcsec$\times$13\arcsec $Ks$-band field
around \Q1429 from ISAAC. North is to the top, East is to the left. b)
Same area,  but with  the PSF  subtracted on components  A and  B (see
Sect.  \ref{sub} for  details).  c) Result  of  the MEM  deconvolution
(Sect. \ref{deconv}): A and B are the two quasar components. d) Result
of the  MCS deconvolution (Sect.  \ref{deconv}). Two very  red objects,
VR1 and VR2 are identified to  the North of component A in addition to
the objects \#1 to 5 (Sect. \ref{vici} and Table \ref{photometry}). }
\end{center}
\end{figure*}

\begin{figure*}
\begin{center}
\caption{\label{H}a)  The   13\arcsec$\times$13\arcsec  $H$-band  field
around \Q1429 from HST/NICMOS2. A and B are the two quasar components. North
is to  the top, East is  to the left. b)  Same area, but  with the PSF
subtracted on  components A and  B (Sect. \ref{sub} for  details). The
faint objects identified in the field are objects previously identified as \#1, 3, 4 and 5 , as
well as three very red objects: VR1  and VR2 to the North of quasar A,
and VR3, to the South (Sect. \ref{vici} and Table \ref{photometry}). }
\end{center}
\end{figure*}

\subsection{PSF-subtraction}\label{sub}

The  determination  of  the  PSF  was
performed  using  DAOPHOT (Stetson  1987).   This
software decomposes the  PSF into the sum of  an analytical profile (a
Moffat profile  in the present case)  and a residual  map.   The analytical profile  and the  residual map  are computed
from as many  stars as possible across the frame  in order to optimize
the  signal-to-noise ratio  of  the PSF.   In  the case  of the  ISAAC
observations, the task is complicated by a slight non-linearity of the
ISAAC detector (1\% at 10000  ADU).  This non-linearity may affect the
two  quasar  components  in   a  different  way,  as  their  magnitude
difference  is  large,  $\Delta$J $\sim\Delta$K$\sim$3.6.   Two  PSFs  were  therefore
constructed, using stars  with a brightness comparable to  that of the
object to be subtracted: a ``faint  PSF'' was built for quasar B (from
2  stars), while  a ``bright  PSF'' was
built for quasar A (from 5 stars).\\
The $R$-band image  is not affected by non-linearity effects.  However, the center
of   \Q1429 A   is   saturated. We   setup   a   special
core-correction  for the PSF  subtraction, giving  more weight  to the
wings of the PSF  than to its core: this is limiting the area over 
which one can trust the PSF subtraction but allows accurate subtraction of the PSF wings.\\
The  PSF determination  for the  HST/NICMOS2 data  was performed  in a
different  way. The  small field-of-view  and the  absence  of stellar
objects close to the quasar components make it impossible to construct
an empirical  PSF.  Instead,  we have used  the TinyTim  V6.0 software
(Krist   \&  Hook   2001)  to   model  the   NICMOS2/F160W   PSF  over
3\arcsec\, $\times$ 3\arcsec.   We have  assumed a  flat spectrum  for the
quasars across  the F160W filter. The  PSF has been  oversampled with 5
$\times$ 5 sub-pixels in order to improve the accuracy of the centering
of the PSF before the  subtraction is carried out.  PSFs were computed
at   the  position  of   the  two   quasar  components,   were  fitted
independently and subtracted from the quasar images.

\subsection{Image deconvolution}\label{deconv}

Two algorithms were used for image deconvolution: the Maximum Entropy
Method  (MEM; Cornwell  \&  Evans, 1985)  and  the MCS  deconvolution
algorithm (Magain et al. 1998). \\
In  the case  of  the MEM  procedure,  we have  deconvolved fields  of
13\arcsec\,$\times$ 13\arcsec  ($R$-, $J$-  and $Ks$-bands), around  the quasar
components. We have used the  PSFs constructed for PSF subtraction and
described above.\\
The PSF used for the MCS deconvolution is similar to that described in
Sect. \ref{sub}, but without  taking into account the non-linearity of
ISAAC.  This  does not  affect  seriously  the  $J$-band  observations
although it introduces  extra residuals close to quasar  A in the $Ks$
filter.  Since quasar A is saturated in the $R$-band, 5 central pixels
of  its   image  were  not   used  in  the  deconvolution.    The  MCS
deconvolution  algorithm  allows  to  choose  the  final  image  pixel
size. We  have adopted for  each band a  pixel size two  times smaller
than  in  the  original  data.   The resolution  after  processing  is
FWHM$_R$=0.30\arcsec\,, FWHM$_J$=0.21\arcsec\,             and
FWHM$_{Ks}$=0.21\arcsec.  The  MCS  deconvolution  algorithm  produces
deconvolved  images and  residual maps  which allow  to  check locally
whether the deconvolved image is compatible with the data
(see Courbin  et al. 1998 for  more details).  Close to  quasar A, the
residuals remain  significant. The size  of the area  with significant
residuals can be  taken as the  size of the zone where  it is not
possible  to detect  a faint  structure.  This  zone has  a  radius of
0.3\arcsec\, in $R$ and $J$, and  a radius of 0.6\arcsec\, in $Ks$, more
affected  by the  non-linearity  of the  detector.   We performed  the
astrometry and  photometry of \Q1429  A and B,  as well as  of the objects
identified on  the deconvolved images.  Results are given in Table
\ref{photometry}.

\begin{table*}[t!]
\renewcommand{\arraystretch}{1.0}
\centering
\begin{center}
\caption[]{Photometry    of    the     objects     in    the
13\arcsec$\times$13\arcsec   deconvolved   field.   The IDs correspond to sources identified  in Figs. \ref{R}, \ref{J}, \ref{K} and \ref{H}. The columns successively provide: (1) the object identification, (2-3) the coordinates relative to A, (4-7) the R, J, H and Ks magnitudes, (8-9) the redshift (spectroscopic redshift  for A and B, photometric redshift for other objects) and the corresponding error bars, (10) some comments or references. The second line provides the date of observation, while the fifth line provides the magnitude differences between quasars A and B.  }
\begin{tabular}{lccccccccc}
\hline
ID & $\Delta \alpha$ & $\Delta \delta$ & R& J & H$_{F160w}$ & Ks& z & $\Delta$z & Comments  \\
   &  (\arcsec)      & (\arcsec)  &  Apr 2001     & Jun 2001   & Dec 1997            & Jun 2001   &            &   &   \\

\hline
\hline
A &0&0&17.45$\pm$0.02&16.32$\pm$0.01&15.88$\pm$0.01&14.96$\pm$0.01&2.076&0.02& Hewett et al. \\
B &-4.455$\pm$0.005 &+2.489$\pm$0.005 &20.85$\pm$0.02& 20.00$\pm$0.01& 19.27$\pm$0.01&18.58$\pm$0.01&2.076&0.02& 1989\\
$\Delta$~m$_{AB}$& & & 3.40$\pm$0.02 & 3.68$\pm$0.01 & 3.39$\pm$0.01&3.62$\pm$0.01 & & & \\
\hline
1&+0.31$\pm$0.05 &-4.50$\pm$0.10&26.08$\pm$0.05&22.66$\pm$0.15&22.18$\pm$0.05&20.08$\pm$0.17&1.1& 1.0-1.3 &$\chi^2>$1 \\
2&-1.50$\pm$0.20&-3.70$\pm$0.20&24.43$\pm$0.05&23.81$\pm$0.20 &$>$ 22.5&21.35$\pm$0.30&\_&\_&\_\\
3& -7.86$\pm$0.01&-1.33$\pm$0.02&24.52$\pm$0.02&21.69$\pm$0.08&20.92$\pm$0.03&20.02$\pm$0.15&0.8&0.6-1.1&\_\\
4&-5.23$\pm$0.04&+5.23$\pm$0.08&22.71$\pm$0.04&21.34$\pm$0.06&21.52$\pm$0.05&20.16$\pm$0.20&1.3&1.0-1.9 &$\chi^2 > $1\\
5&-1.22$\pm$0.14&+4.11$\pm$0.02&23.53$\pm$0.06&21.66$\pm$0.07&21.17$\pm$0.04&20.30$\pm$0.15&1.2&0.8-1.9&\_\\
6&+0.82$\pm$0.10&+0.96$\pm$0.10&24.89$\pm$0.06&$>$ 24.5&$>$ 22.5&$>$ 24.5&\_&\_&\_\\
7&-0.55$\pm$0.10&+1.43$\pm$0.10&23.18$\pm$0.02&$>$ 24.5&$>$ 22.5&$>$ 24.5&\_&\_&\_\\
\hline
VR1&+1.02$\pm$0.10&+2.01$\pm$0.10&$>$27&$>$24.5&19.30$\pm$0.01&20.49$\pm$0.15&\_&\_&\_\\
VR2&+0.26$\pm$0.10&+2.36$\pm$0.10&$>$27&$>$24.5&21.69$\pm$0.04&21.60$\pm$0.40&\_&\_&\_\\
VR3&+1.60$\pm$0.10&-2.6$\pm$0.10&$>$27&$>$24.5&20.99$\pm$0.03&$>$24.5&\_&\_&\_\\
\hline
\label{photometry}
\end{tabular}
\end{center}
\end{table*}

\subsection{Results of the PSF subtraction and of deconvolution}\label{res}

\begin{table}[t!]
\renewcommand{\arraystretch}{1.0}
\centering
\begin{center}
\caption[]{Lower  limit  magnitude  of detectability of a putative lensing galaxy 
according to its  position (column 2, 3 and 4). Column 5 displays the
minimal redshift for an $\mathcal{L}^*$ galaxy of this magnitude. }
\begin{tabular}{lcccl}
\hline
Position  & R& J & Ks&   \\
\hline
\hline
(a) 0.8\arcsec\, East to A  & 25.5 & 24.0    &22.5 &z$>$1.9\\
(b) 0.4\arcsec\, West to B&  26.5 &24.5    &24.5 &z$>$3.2\\
(c) 7.0\arcsec\, South-West to A  &27.0 &24.5 & 24.5          &z$>$4.5 \\
\hline
\label{limit_mag}
\end{tabular}
\end{center}
\end{table}

Results of  the PSF  subtraction are  displayed on Figs.  1 to  4.  We
estimate that  sources could not be detected if they were located
within radii of 1.0\arcsec\,($R$), 0.5\arcsec\,($J$), 0.7\arcsec\,($H$)
and 0.6\arcsec\,($Ks$) from component A, because of the difficulty to build a PSF as bright as quasar A. On the contrary, for quasar B,
faint sources could be detected after PSF subtraction, without spatial
limitation.   Some residual  signal above  the sky  background  can be
detected after the PSF subtraction of components A and B, at the level
of 1.8$\sigma$ in $Ks$ and  2.0$\sigma$ in $J$.  It is unclear whether
these extended residuals are artifacts  of the PSF subtraction or real
faint extended sources.

In order to  set the  limits of this detection,  we have added a
fake  elliptical  galaxy of a size comparable to that of galaxy \#1 found in the environment of the quasars  (major  axis a=0.8\arcsec,  minor  axis
b=0.4\arcsec). This simulation has been made on the  $R$, $J$  and $Ks$ images, placing the fake  at  various positions
across  the  image:  (a) 0.8\arcsec\, East   to  the  center  of  quasar  A, (b)
0.4\arcsec\, West to the  center of quasar B (these  positions are within
the wings of the quasar  images), (c) 7\arcsec\, South-West to the center of
quasar A, where  no signal had been detected  after PSF subtraction or
deconvolution.  These simulated frames  have been deconvolved using the
MEM   procedure,   in a way similar to   that   described   in
Sect. \ref{deconv}. We increased the magnitude of the fake  by steps of 0.5 from R=24, J=21 and Ks=20 up to  the magnitudes at which a
galaxy could  not be detected above the noise (S/N=1). These limiting  magnitudes are displayed in Table \ref{limit_mag}.  Comparing to the apparent  magnitude of an elliptical galaxy with
M$^{\star}_{B_{T}}$=-20.59  (Cheng   \&  Krauss  2001)   at  different
redshifts, this simulation  allows  to infer  the  maximum redshift
accessible in our processed dataset, as a function
of  the  putative  lens  position. 
In the  case of  a lensing galaxy  located between  the two
quasars (position c), or very  close to  quasar B (position b),  we would easily  probe until the
quasar redshift (z$_{quasar}$=2.076)  after deconvolution. We conclude
that no galaxy with $\mathcal{L}\sim\mathcal{L}^{\star}$  is detected in this area. If
the  putative lens  were located  on the  wings of  quasar A (position a),  we could
detect it up to  z=1.9 (see Table \ref{limit_mag}) . Such a  source would be too close to the quasar
itself to act as a lensing object.
 
\section{The quasars and their immediate vicinity }\label{vici}

The magnitude differences between the two quasars in the four bands are  displayed in Table \ref{photometry}. They vary between 3.4 and 3.7 mag, depending on the band.  In the case of gravitational lensing, this value should in principle  be constant, whatever the wavelength is. However, such small variations may arise from various effects. For example, the intrinsic temporal variability of the quasars could explain the lowest  $\Delta$m$_{AB}$ in the $H$-band,  taken 4 years before the $J$ and $Ks$ datasets (similar case for the quasar SBS~1520+530 studied by  Faure et al. (2002)). Also, the $R$-band image is  expected to be more  affected by absorption than the near-IR images. As the quasars are separated by 5.1\arcsec, the different $\Delta$m$_{AB}$ measured could probe the presence of  different absorbers along the line of sight towards the quasars A and B. Finally, microlensing effects could also explain these small differences ($<$0.3 mag) between the different bands. In conclusion, the variation with wavelength of the  intensity ratio between the two quasars does not allow to reject the hypothesis of a lens for  this system.\\

Objects \#1, 3, 4 and 5 being detected in the 4 bands,  we can tentatively  estimate
their photometric redshifts. We use the \hyp  software,  a  SED fitting procedure  based on the fit  of the
overall shape  of the  galaxy spectra and  on the detection  of prominent
spectral features (Bolzonella  et al. 2000). The observed photometric SED is compared to those
obtained from  a set of reference spectra  (spectral evolution library
of Bruzual \&  Charlot, 1993), using the same  photometric system. The
observed  galaxy  flux  is  corrected  for  reddening  following  the
Calzetti  law  (Calzetti et  al.  2000).   \hyp  performs a  $\chi^2$
minimization and  the redshift given in  Table \ref{photometry} corresponds
to  the lowest  $\chi^2$ (ideally  $\chi^2  < $1),   while the  photometric
redshift  error  bars  are  the  extrema of the confidence  interval  at
68\%. Our set of filters  allows to bracket  the 4000\AA~  break from
z$\geq$0.5. Yet, it has been shown   that 4  bands is still too  few to derive accurate photometric redshifts (Bolzonella et al. 2000, Athreya et
al. 2002). With this remark in mind, we find a mean redshift of  z$\sim$1 for objects \#1, 3, 4 and 5 (with error bars up to 50\%), the SED of objects \#1, 3 and 5 beeing best fitted by a starburst type galaxy, and that of object \#5, by an Im type galaxy. Object \#2,  which is not detected in the $H$-band, shows  magnitudes in the other bands  comparable  to those  of  galaxies \#1, 3, 4  and 5.

\begin{figure*}[!th]
\begin{center}
\includegraphics[width=8.3cm]{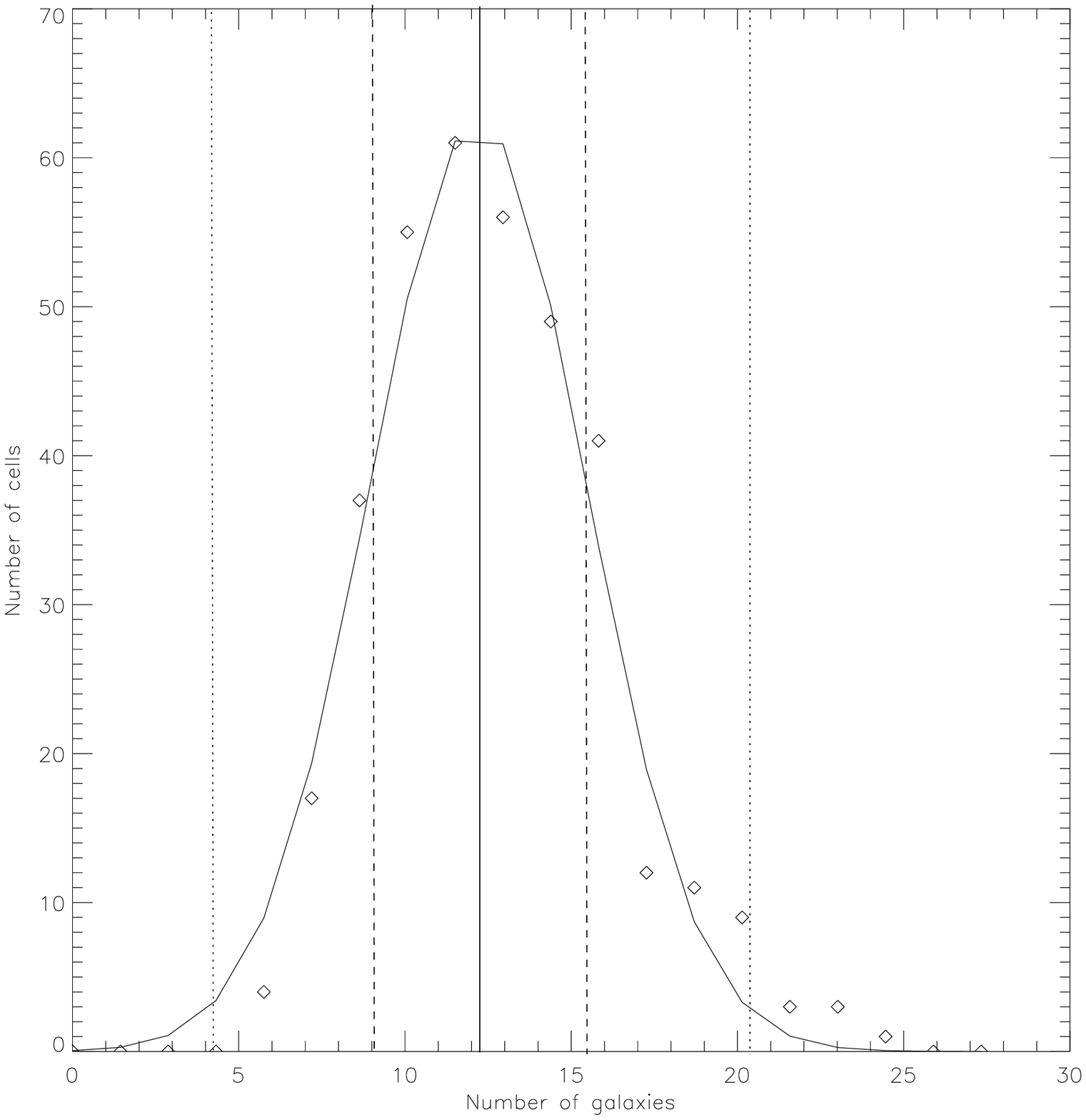}
\caption{\label{iso} The  left panel  (see 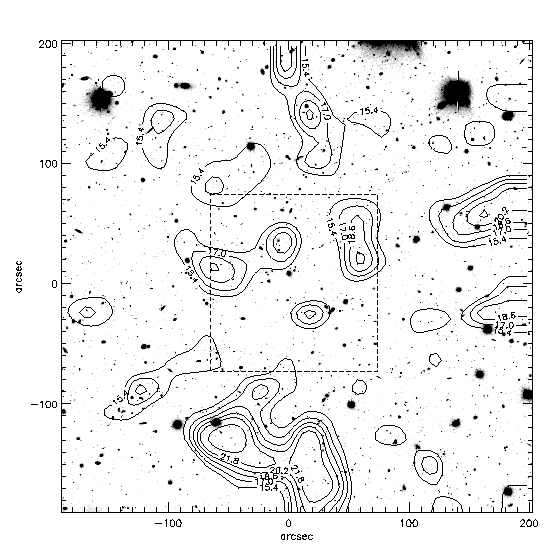) displays the number density  isocontours  for the
magnitude range 23$\leq$  R $<$25, superimposed on the  FORS1 image. The
coordinate (0,0)  indicates the location of \Q1429 A. The contours
outline the regions of constant density. Labels of the contours display the
number  of galaxies  per arcmin$^2$  (starting at 1$\sigma$, and then,
1.5$\sigma$,  2$\sigma$, 2.5$\sigma$,  3$\sigma$  and 4$\sigma$).  The
dashed square represents the ISAAC  field-of-view. North is to the top,
East to  the left. The right panel  shows the number of  cells in the
field which have a given galaxy density (galaxies per arcmin$^2$). 
The Gaussian fits
the points corresponding to cells which are assumed to be filled with background galaxies. The solid  vertical line displays the mean  galaxy density of
the distribution, whereas the dashed and dotted lines are respectively
at 1$\sigma$ and 2.5$\sigma$.}
\end{center}
\end{figure*}

Two objects, close to quasar A, are  detected only in the $R$-band.
They are labeled \#6 and 7  in Table \ref{photometry} and may be in fact
one  single object, artificially separated  by the  deconvolution and  PSF
subtraction procedures at the position  of quasar A.  These
objects could pertain to the image  of the amplified and distorted quasar host
galaxy. In this  case, we would expect a counter-image of object \#6
in the B component of the quasar, to have R$\sim$28.3, according to the flux ratio measured between quasars A and B. The counter-image
of object \#7 would be expected at R$\sim$ 26.6.  This  is at  the limit  of the
data, and although some very faint, non significant residuals are seen
close  to image B  (see  Fig.   \ref{R}) we  cannot report on  any  firm
detection of the quasar host counter-image in component B. Given our detection limits in the near-IR,
and given the colors of a typical quasar host, we do not expect to see
any trace of a near-IR counterpart of objects \#6 and \#7 in our ISAAC 
observations. Our observation in the $R$-band is therefore consistent with
the detection of the quasar host, but it could well be also a relatively blue object at low redshift.

Finally, 3  very red objects  (VR1, VR2 and  VR3 in Fig.  \ref{H}) only
visible in the  $H$- and $Ks$-bands, are seen close  to quasar A. VR1 and VR2
have R-K$>$5.3 and thus are belonging to the ERO class of objects (e.g. Thompson et al. 1999). VR2 is at the detection limit of the VLT data, but its position for  flux measurement was inferred from the HST data. VR3 is only detected in $H$ with R-H$>$6. If these objects are at the redshift
of the quasar, the 4000\AA~ break falls between the $J$- and $H$-bands and
would explain the observed red colors. Alternatively, they could be obscured star forming galaxies at redshift $\sim$1 (e.g.  Spinrad et al. 1997, Cohen et al. 1999, Smith et al. 2002a and 2002b).
 In that case, they
might have redshifts compatible with  those of objects \#1, 3, 4 and 5
and be faint members of a galaxy group at z$\sim$1.

\section{Looking for a cluster lens}

The spatial  distribution of objects in  the field of  \Q1429 has
been investigated. Object catalogs have been  constructed in the
$R$-,  $J$-, $H$-  and $Ks$-bands,  using the  SExtractor 1.2  software
(Bertin \&  Arnouts 1996).  The  separation between  stars  and galaxies
was made using a combination of the FWHM versus magnitude plot and of the
analysis    of    the     peak    surface    brightness    $\mu_{max}$
(mag~arcsec$^{-2}$)  of  the  objects  versus their  total  magnitude
(Bardeau et al.  2003). As a matter of fact, non-saturated stars are bright point-like objects of the size of the PSF. Therefore, in the FWHM versus magnitude plot, they are easily detectable on an horizontal line. Moreover, the stars have the highest peak surface brightness, and  for stars, $\mu_{max}$ increases linearly with magnitude. Therefore, using these two plots it is possible to separate galaxies from stars in our catalogs of objects. Finally, our photometric catalogs are composed of $\sim$3000
galaxies in $R$, and  $\sim$250  in $J$ and $Ks$. 
 
\subsection{Isodensity contours}

We have used the 6.8\arcmin$\times$6.8\arcmin\,  $R$-band image to compute the
isodensity  contours of  the galaxy  distribution. The galaxy
catalog has been split in  three magnitude  bins (22$\leq$R$<$24, 23$\leq $R$<$25,
24$\leq $R$ <$27) that allow  to study roughly  3 redshift slices.  As we
do not  have any  reference-field  of the same  depth for  comparison, we have
determined the  mean background galaxy density from the data itself,  dividing the
image into a grid of  20$\times$20 contiguous cells of 20\arcsec\, side.
The cell size corresponds to  50 kpc  at z=1.   The
galaxy  density  is  measured  in   each  cell  and  an  histogram  is
constructed of the frequency of occurrence of each density (right panel
of Fig. \ref{iso} for the magnitude range 23$\leq$ R$ <$25).  A Gaussian
is fitted to the histogram,  giving a mean density and a standard
deviation  which will be used in estimating the  significance  of an eventual   cluster detection. A 2.5$\sigma$ deviation (20.2 galaxy per arcmin$^2$ in
this plot) away  from the mean  can be considered  as a significant galaxy overdensity.

 The
distribution of  bright galaxies (22$\leq$R$  <$24)   does not show   any
significant  overdensity. Taking into account all  very faint  objects down  to our
limiting magnitude (24$\leq$ R$ <$27)  does not allow either to unveil any
overdensity.  Cutting the faint end of our galaxy catalog (23$\leq$R$<$25),  i.e., removing  possible spurious  detections, results  in one
 peak  located 2.3\arcmin\,South-East to \Q1429A. (left
panel of Fig.  \ref{iso}). This overdensity is  too far away from the quasars  to be the source of any lensing  effect.  We have performed a
similar analysis for the  $J$ and $Ks$ images,  and do not
find any galaxy overdensity above 2.5$\sigma$.

\begin{figure}[!th]
\begin{center}
\includegraphics[width=8.5cm]{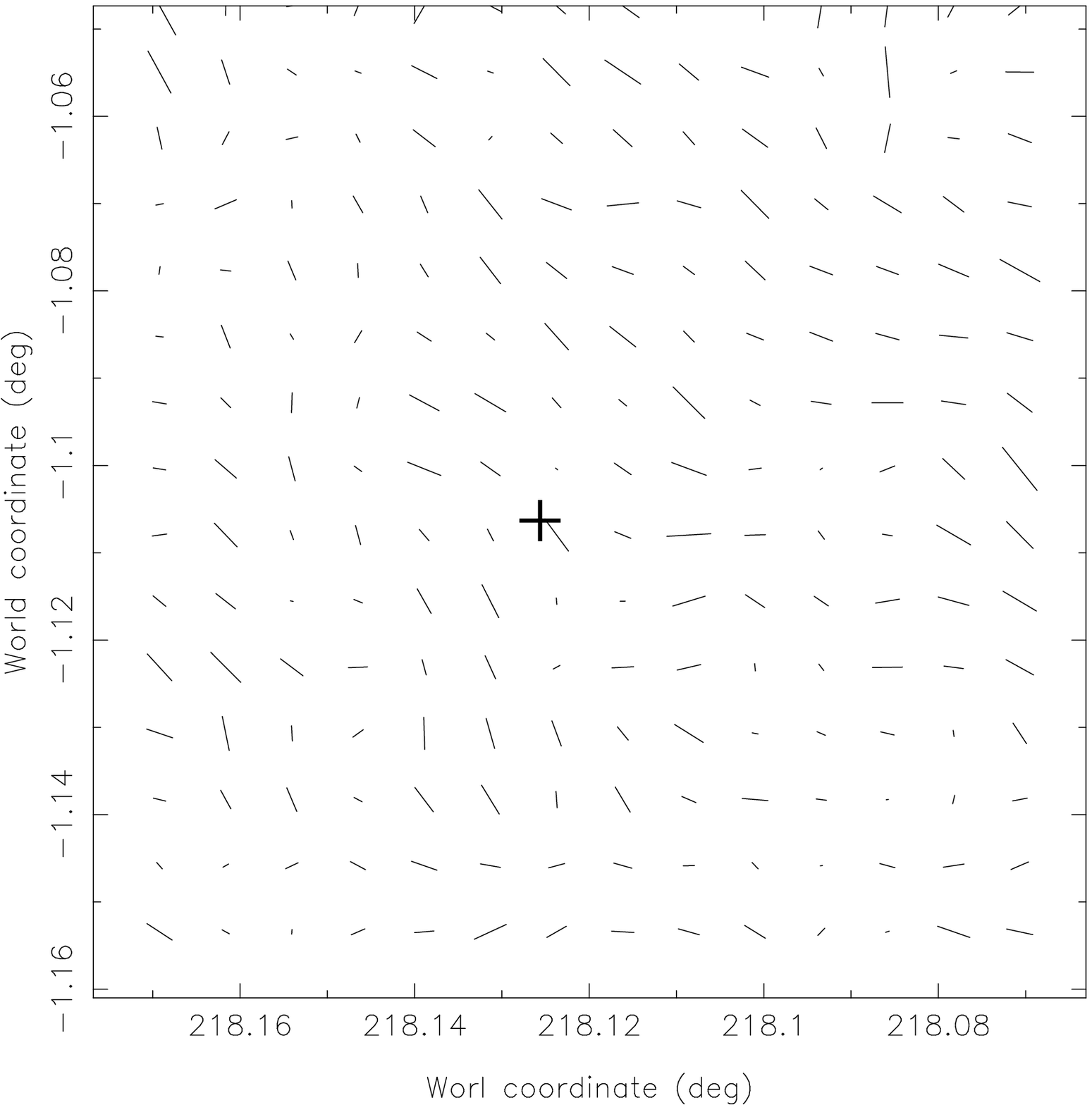}
\caption{\label{shear} The  top  panel (see 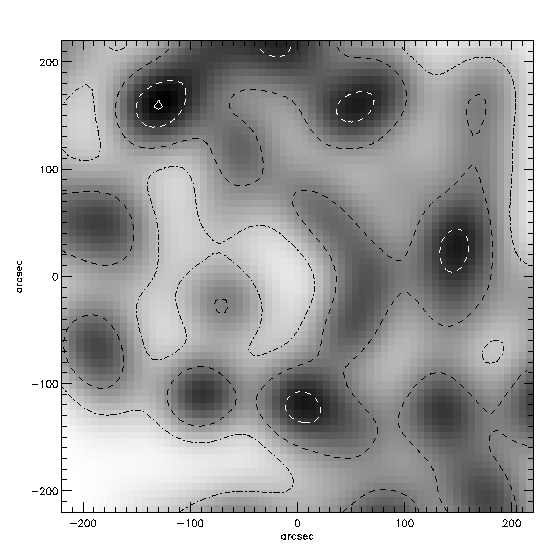) shows  the  shear  map for  the
6.8\arcmin\,$\times$6.8\arcmin\,  field-of-view around  the  quasars. Each
stick  displays  the  mean  direction  of  galaxies  in  a  cell  of
30\arcsec\,$\times$30\arcsec ($\epsilon_{max}$=0.23) .   The    coordinates   (218.125,   -1.105)
indicates the position  of \Q1429 A (black cross).  The bottom panel  shows the surface
mass density from the maximum-entropy mass reconstruction. On this grey-scale,white corresponds to the lowest mass level, and  black corresponds to the highest mass level. The contours successively outline regions of signal to noise: 0.3 (black dotted-dashed line), 0.6 (black dashed line), 1 (white dashed line) and 1.2 (white continuous line), the noise level being computed from the bootstrap sample.}
\end{center}
\end{figure}

\subsection{Weak-shear analysis}

If \Q1429 is lensed by a group of galaxies or a cluster of  galaxies (dark or visible), a weak lensing analysis
of  the  field  should reveal  it.   The  best available data  to carry  out  this
experiment is the $R$-band image, which is  very deep (limiting magnitude in R=27 mag) and has the largest field-of-view (6.8\arcmin\, $\times$ 6.8\arcmin).

The  statistical  distortions  of  field  galaxies by  a galaxy
cluster, can be  quantified by measuring the mean  ellipticity of many
faint galaxies in  cells of a given size.  This size  has to be chosen
carefully: if it is much  larger than the scale of  an appreciable variation of
the  potential,  the  averaging  will  destroy the  coherence  of  the
distortion pattern, while if the cells  are too small, the  averaging will
not  remove the  effect  of galaxy  intrinsic  ellipticities from  the
distortion estimation and the result  will be dominated by shot-noise.

For any weak-lensing signal to be accurately  measured, the first  step  is to correct the  observed galaxy ellipticities for
the  instrumental distortions.  To do  this,  the  PSF variation  is
measured across the field from 175 stars, using the {\it  Im2shape} software (Bridle
et al.  2002).  The ellipticity of the PSF varies from $\epsilon$=(a-b)/(a+b)=0.009 to 0.034 across the FORS1 field.  Each  background galaxy (R$\geq$23 mag) is deconvolved in an analytical way, modeling it by an adequate 2D-function and   using the  closest  PSF. The  corrected  galaxy ellipticities  are
averaged   in   cells   of  20\arcsec\,$\times$20\arcsec\,   size,   which
  allow to use 10 to  50 galaxies by
cell. The mean ellipticity of the galaxies vary from  $\epsilon$=(a-b)/(a+b)=0.05 to 0.23 across the image.  In
a field  revealing weak shear,  we expect the mean  ellipticities from
one cell to  the other to be correlated.  
 In Fig. \ref{shear} (top panel)  we display the shear-map, where each stick orientation indicates the mean galaxy orientation in the  cell, and where the length of the stick is proportional to the mean  ellipticity $\epsilon$. The mean ellipticity over the FORS field computed as ($<$e$_1>$,$<$e$_2>$)\footnote{$<$e$_1>$= $<\epsilon_i\times$cos(2$\times \theta_i$)$>$ and   $<$e$_2>$= $<\epsilon_i\times$sin(2$\times \theta_i$)$>$, for each stick $i$}  is (-0.04+/-0.03, 0.00+/-0.04). It does indicate a slight preferred orientation along the horizontal axis, but this is consistent with the typical cosmic variance. Thus, there is no evidence that a strong lens is present in the vicinity of this field. The weak shear analysis of the $J$ and $Ks$
frames reveals features similar to those in the $R$-band, and therefore does not show either any coherent distribution of the mean galaxy ellipticities revealing the presence of a galaxy cluster in our field-of-view.

We  used  the  maximum-entropy  algorithm {\it  LensENT2}  (Bridle  et
al. 1998, Marshall et al. 2002)  to study  the spatial distribution  of the mass.   The lens
plane is assumed  to be at z=1 (corresponding to  the mean redshift of the
closest   faint   galaxies   detected   near   quasar   A,   see
Sect. \ref{vici}). The weak shear  has been computed for galaxies with
 R $\geq$  23.  The mass profile or Intrinsic Correlation Function (ICF) of the mass clumps is assumed to be Gaussian. The width of  the ICF, that
defines the final spatial resolution of the map, is not intrinsically constrained due
to the  low coherence in  the shear pattern.  An ICF of  60\arcsec\, was
therefore  assumed.  The  reconstructed  mass  map  is  shown  in  Fig.
\ref{shear} (bottom panel).

Since the  noise properties  of maximum-entropy inversion  methods are
not easily interpretable, we have performed 100 reconstructions on modified
catalogs of  galaxies for which  positions are retained  but ellipticity
directions  are assigned  at  random.  We have compared  the  level of  the
structures detected in the  reconstructed map to the bootstrap sample:
the highest peak (2.8\arcmin\,, North-East to \Q1429 A) in the map reconstructed from our dataset is 1.2 times higher than the highest peak in the bootstrap sample. Thus, our detection is very close to the noise level. 

If it were actually a signal, let us quantify it in term of mass.
 We performed the following simulation: assuming that the field galaxies above R=23 mag are background galaxies with a mean redshift of z=1.5, we have  added a Spherical
 Isothermal Sphere (SIS) at z=1 in the line of sight towards this field, and we have computed its influence on the background galaxies. The resulting fake frame has been analyzed in a way similar to  the original frame: we performed a weak-shear analysis and a mass reconstruction. To discard the risk of having superimposed the SIS on top of a possible faint and undetected mass potential, we did the simulation for several SIS 
positions across the field-of-view: 2\arcmin\, South-East, 2\arcmin\, South-West, 2\arcmin\, North-West and 2\arcmin\, North-East to \Q1429 A, as well as at the position of \Q1429 A. We have varied the mass of the potential until we could detect clumps in the reconstructed mass map of the same level as in the original data. 
This mass varies between 1.2~10$^{13}\mathcal{M}\odot$~arcmin$^{-1}$ and  2.5~10$^{13}\mathcal{M}\odot$~arcmin$^{-1}$ 
(from 3 to 6~10$^{13}\mathcal{M}\odot$ for a lens at z=1), or, in terms of velocity dispersion, between $\sigma_v$=320 and 500 km~s$^{-1}$. 
This is the limit to the detection of weak lensing in this field using the current data.

\section{Conclusions}

Using deep VLT and HST images,  we have searched, on several scales, for
a gravitational lens in the field of the quasar pair  \Q1429 A and B.

On the scale of galaxy cluster, we  do not detect  any significant galaxy 
overdensity     in    a   6.8\arcmin\,$\times$6.8\arcmin\,
field around the quasar pair, down to R$\sim27$. Under the hypothesis
of a  dark lens, we have  performed a weak  lensing analysis. Although
our  technique would allow to  detect mass  concentrations down to
1.2-2.5~10$^{13}\mathcal{M}\odot$~arcmin$^{-1}$,  we  do not  detect  any coherent  shear pattern on faint field galaxies. We conclude that there is no evidence
for a large scale lens in the field of \Q1429,  whether it
be luminous or  dark.

On smaller  scales ($\sim$1-5\arcsec), no luminous  object is detected
between the quasar images down  to our limiting magnitudes in the four
filters (27~mag in $R$, 24~mag in $J$ and 22.5~mag in $Ks$) .  If  the lens in \Q1429  were within the  usual redshift range
observed   for   lens  galaxies   (0$<$z$<$1),   simple  (and   vastly
under constrained)  modeling with  an isothermal  sphere shows  that the
velocity dispersion  of the lens  should be of  the order of  350 km~s$^{-1}$. According to the Faber-Jackson relation, this implies a galaxy 6 times
brighter  than an $\mathcal {L}_{\star}$ galaxy ($ \mathcal{M}/\mathcal{L}_{J}\sim$3000),  which  would be easily detected  given  the
depth of the available dataset. Tentative modeling of a multiple  lens has been carried out, involving
objects  \#1 to 5  in the  immediate vicinity  of the  quasar pair. These
objects   have  similar  colors,   compatible  with   ellipticals  at
z$\sim$1.  We   failed however  in  reproducing both the  image
separation and flux  ratios, using such a model. The group of 5 galaxies cannot be alone responsible for the \Q1429 A and B configuration, would it be a lensed quasar.

Finally, the only  remaining sign of lensing is  the detection of what
might be the lensed host galaxy of \Q1429, detected in the $R$-band, as
objects  \#6 and  7.  Given  the   limiting magnitudes  in the  other
filters, we do not expect to detect it in the near-IR, nor do we expect to see
its  counter-image in  the  B component  of  \Q1429. Another  possibility is that this
object is  a foreground blue object, unrelated to the quasar. Two  observations  might settle  the issue. First,  a deep  HST image
with R$>$27, as could be  obtained with the Advanced Camera for Surveys
should allow  to constrain the morphology  of objects \#6  and 7 well
enough, both for  quasar A and their counter-part for quasar B, to decide  whether it  is the quasar host galaxy 
stretched by lensing or whether it  is a foreground object. 

Second, the photometric monitoring and the measurement of a time delay
would establish the lensed nature of \Q1429.
 The  first
observation is straightforward and would settle the issue straight. The second observation is more telescope time consuming. 

With  the present observational  material, we  conclude that  there is
very little evidence that \Q1429 is lensed, either  by a galaxy cluster, or a galaxy group or a
single galaxy (dark or luminous) and thus, should be considered as a genuine binary quasar.

\begin{acknowledgements}
We acknowledge the guidance  of Andreas Jaunsen in the  NICMOS2 data reduction.
Micol Bolzonella  provided help in  the application of the Hyperz software to the
present dataset. We are indebted to an anonymous referee for interesting comments.  The HST data used  in this paper were  obtained by the
``CfA  Arizona Space Telescope  LEns Survey''  (CASTLES) collaboration
(PI:  E.  Falco).   C\'ecile Faure  acknowledges support  from  an ESO
studentship in  Santiago.  Fr\'ed\'eric Courbin is  supported by Marie
Curie  grant  MCFI-2001-00242.   The collaborative  grant  ECOS/CONICYT
CU00U005  between Chile  and France  is also  gratefully acknowledged.
Jean-Paul  Kneib thanks  CNRS for  support  as well  as the  ESO Chile
visitor program.
\end{acknowledgements}

\end{document}